\documentclass[pra,twocolumn,floatfix]{revtex4}

\usepackage{amsmath}
\usepackage{graphicx}
\usepackage{amsfonts}
\usepackage{amssymb}

\newtheorem{mytheorem}{Theorem}

\newtheorem{mycorollary}{Corollary}

\begin{document}

\title{Conditions for strictly purity-decreasing quantum Markovian dynamics}
\author{D.A. Lidar$^{(1)}$, A. Shabani$^{(2)}$, and R. Alicki$^{(3)}$}
\affiliation{$^{(1)}$Chemical Physics Theory Group, Chemistry Department, and Center for
Quantum Information and Quantum Control, University of Toronto, 80 St.
George St., Toronto, Ontario M5S 3H6, Canada\\
$^{(2)}$Physics Department, and Center for Quantum Information and Quantum
Control, University of Toronto, 60 St. George St., Toronto, Ontario M5S 1A7,
Canada\\
$^{(3)}$Institute of Theoretical Physics and Astrophysics University of Gd\'ansk, Poland}

\begin{abstract}
The purity, $\mathrm{Tr}(\rho ^{2})$, measures how pure or mixed a
quantum state $\rho $ is. It is well known that
quantum dynamical semigroups that preserve the identity operator
(which we refer to as unital) are strictly purity-decreasing
transformations. Here we provide an almost complete characterization
of the class of strictly purity-decreasing quantum dynamical
semigroups. We show that in the case of finite-dimensional Hilbert
spaces a dynamical semigroup is strictly purity-decreasing if and only
if it is unital, while in the infinite dimensional case,
unitality is only sufficient.

\end{abstract}

\maketitle

\section{Introduction}

\label{intro}

Quantum dynamical semigroups have been studied intensely in
the mathematical and chemical physics literature since the pioneering
work of Gorini, Kossakowski and Sudarshan \cite{Gorini:76} and
Lindblad \cite{Lindblad:76}. They have a vast array of applications,
spanning, e.g., quantum optics, molecular dynamics, condensed matter,
and most recently quantum information
\cite{Alicki:87,Gardiner:book,Breuer:book,Nielsen:book}.

In this work we are interested in general conditions for dissipativity
\cite{Lindblad:76}, namely the question of which class of quantum dynamical
semigroups is guaranteed to reduce the purity of an arbitrary
$d$-dimensional state $\rho $, where the purity $p$ is defined
as
\begin{equation}
p=\mathrm{Tr}\rho ^{2}\text{.}
\end{equation}
The purity, which is closely related the Renyi entropy of order $2$,
$-\mathrm{\log Tr}\rho ^{2}$ \cite{Ingarden:book}, satisfies
$1/d^{2}\leq p\leq 1$, with the two extremes $p=1/d^{2},1$
corresponding, respectively, to a fully mixed state and a pure
state. The question we pose and answer in this work is:

\begin{quote}
What are the necessary and sufficient conditions on quantum dynamical semigroups for the purity to be
monotonically decreasing ($\dot{p}\leq 0$)?
\end{quote}

To answer this question we first revisit a general expression for
$\dot{p}$ using the Lindblad equation, in Section~\ref{Lind}. We then
give a derivation of a sufficient condition for purity-decreasing
quantum dynamical semigroups in Section~\ref{suff}. This condition is valid for
a large class of, even unbounded, Lindblad operators. We establish a necessary
condition in Section \ref{nece}, which is valid in finite-dimensional Hilbert spaces. Some examples are
also discussed in this section. Concluding remarks are presented in
Section~\ref{conclusions}.

\section{Purity and Markovian Dynamics}

\label{Lind}

The action of a quantum dynamical semigroup can always be
represented as a master equation of the following form (we set $\hbar =1$): 
\begin{equation}
\frac{\partial \rho }{\partial t}=-i[H,\rho ]+{\mathcal{L}}(\rho ),
\label{eq:Lind}
\end{equation}
where $H$ is the effective system Hamiltonian, and the Lindblad
generator ${\cal L}$ is
\begin{equation}
{\mathcal{L}}(\rho )=\frac{1}{2}\sum_{\alpha ,\beta }a_{\alpha \beta }\left(
[G_{\alpha },\rho G_{\beta }^{\dagger }]+[G_{\alpha }\rho ,G_{\beta
  }^{\dagger }]\right) .
\label{lind}
\end{equation}

The matrix $A=(a_{\alpha \beta })$ is positive semidefinite [ensuring
complete positivity of the mapping ${\mathcal{L}}(\rho )$], and the Lindblad
operators $\{G_{\alpha }\}$ are the coupling operators of the system to the
bath \cite{Alicki:87}. One can always diagonalize $A$ using a unitary
transformation $W=(w_{\alpha \beta })$ and define new Lindblad operators $
F_{\alpha }=\sqrt{\gamma _{\alpha }}\sum_{\beta }w_{\alpha \beta }G_{\beta }$
such that 
\begin{equation}
{\mathcal{L}}(\rho )=\frac{1}{2}\sum_{\alpha }\left( [F_{\alpha },\rho
F_{\alpha }^{\dagger }]+[F_{\alpha }\rho ,F_{\alpha }^{\dagger }]\right) ,
\label{eq:lind2}
\end{equation}
where $\gamma _{\alpha }\geq 0$ are the eigenvalues of $A$. Note that, formally, for any $A$
\begin{equation}
\mathrm{Tr}[{\mathcal{L}}(A)]=0 .
\end{equation}
For the bounded semigroup-operators case (hence in particular in
finite-dimensional Hilbert spaces) (\ref{lind}) provides the most
general form of the completely positive trace preserving
semigroup. However, the formal expression (\ref{eq:Lind}) makes sense
with unbounded $H$ and $F_{\alpha}$ provided certain technical
conditions are satisfied \cite{Davies:77}. In particular, $D= -iH
- \sum F_{\alpha}^{\dagger}F_{\alpha}$ should generate a contracting
semigroup of the Hilbert space and $F_{\alpha}$ should be well-defined
on the domain of $D$.
Under those technical conditions one can
construct the so-called {\em minimal solution} of (\ref{eq:Lind})
which may not be trace-preserving \cite{Davies:77}. One should notice that in the
unbounded case the solution $\rho(t)$ of (\ref{eq:Lind}) need not
be differentiable unless $\rho(0)$ is in the domain of $-i [H,.]+{\cal L}$.

In the following, when dealing with unbounded generators we tacitly
assume that all these technical conditions are satisfied and then the
formal mathematical expressions can be precisely defined. In
particular, for unbounded $A$, the positivity condition $A\geq 0$
means that $\langle \phi| A | \phi \rangle \geq 0$  for all $\phi$ from a proper dense domain.

The time-evolution of the purity can thus be expressed as
\begin{eqnarray}
\dot{p} &=&\mathrm{Tr}(\rho \frac{\partial \rho }{\partial t})+\mathrm{Tr}(
\frac{\partial \rho }{\partial t}\rho )  \notag \\
&=&-i\mathrm{Tr}(\rho \lbrack H,\rho ])-i\mathrm{Tr}([H,\rho ]\rho )
  \notag \\
&& + \mathrm{
Tr}(\rho {\mathcal{L}}(\rho ))+\mathrm{Tr}({\mathcal{L}}(\rho )\rho ).
\end{eqnarray}
Recall that if $A$ is bounded and $B$ is trace-class then $AB$ and $BA$ are
also trace-class, and $\mathrm{Tr}(AB)=\mathrm{Tr}(BA)$ \cite{Richtmyer:book}.
Assuming bounded
$\{F_{\alpha }\}$ we thus have   
\begin{eqnarray}
\dot{p} &=&-2i\underbrace{\mathrm{Tr}(\rho \lbrack H,\rho ])}_{=0}+2\mathrm{
Tr}(\rho {\mathcal{L}}(\rho ))  \notag \\
&=&\sum_{\alpha }\mathrm{Tr}[\rho F_{\alpha }\rho F_{\alpha }^{\dagger }]-
\mathrm{Tr}[\rho ^{2}F_{\alpha }^{\dagger }F_{\alpha }],  \label{eq:pdot1}
\end{eqnarray}
Thus Hamiltonian
control alone cannot change the first derivative of
the purity, and hence cannot keep it at its initial value (the
\textquotedblleft no-cooling principle\textquotedblright\
\cite{Ketterle:92}); the situation is different when one exploits the
interplay between control and dissipation \cite{Tannor:99}, or with
feedback \cite{LidarSchneider:04}.

It is well-known that a sufficient condition for the purity to be a
monotonically decreasing function under the Markovian dynamics, is
that the Lindblad generator is \emph{unital}, namely ${\mathcal{L}}(I)=0$, where $I$
is the identity operator \cite{Streater:book}. But can this condition be sharpened?
Indeed, we will show that unital Lindblad generators are in fact a special case of
a more general class of quantum dynamical semigroups for which purity is strictly
decreasing. In Section~\ref{suff} we prove the following theorem:

\begin{mytheorem}
  \label{th:1}
A \emph{sufficient} condition for $\dot{p}\leq 0$ under Markovian
dynamics, Eqs.~(\ref{eq:Lind}),(\ref{eq:lind2}), is:
\begin{equation}
{\mathcal{L}}(I)=\sum_{\alpha }[F_{\alpha },F_{\alpha }^{\dagger }]\leq 0.
\label{eq:condition}
\end{equation}
whenever the generally unbounded formal operator (\ref{eq:condition}) can be defined as a form
on a suitable dense domain.
\end{mytheorem}

Note that Theorem~\ref{th:1} places no restriction on Hilbert-space
dimensionality. Moreover, for particular cases condition~(\ref{eq:condition}) is meaningful for unbounded
generators provided certain technical conditions concerning domain
problems are satisfied (e.g., the
amplitude raising channel, discussed below).
Theorem~\ref{th:1} can be sharpened under an additional assumption:

\begin{mytheorem}
  \label{th:2}
  In the case of \emph{finite-dimensional} Hilbert spaces the purity
is monotonically decreasing if and only if the Lindblad generator is
unital,
\begin{equation}
{\mathcal{L}}(I)=\sum_{\alpha }[F_{\alpha },F_{\alpha }^{\dagger }]=0.
\end{equation}
\end{mytheorem}

This is proved in Section~\ref{nece}. Note that in the case of a
finite-dimensional Hilbert space the Lindblad operators are
automatically bounded.

\section{Sufficiency}

\label{suff}

In this section we present three different proofs of sufficiency. The
first is the most general, in that it is valid also for unbounded
operators, under appropriate restrictions. The second and third are
valid only for bounded operators and are presented for their pedagogical value.

\subsection{General proof of sufficiency}

We present a proof of Theorem~\ref{th:1} which is valid even for
unbounded $F_{\alpha}$, under the following technical conditions,
which are satisfied for important examples of Lindblad generators \cite{comment-conds}:\\
\noindent a) There exists a dense subset ${\cal D} \subset $ \{joint
dense domain of all of $\{H, F_{\alpha}, F_{\alpha}^{\dagger}, \sum_\alpha F_{\alpha}^{\dagger} F_{\alpha},
\sum_\alpha F_{\alpha} F_{\alpha}^{\dagger}\}$\};\\
\noindent b) All finite range operators $\sum |\psi_k\rangle
\langle\phi_k|$, where $|\psi_k\rangle, |\phi_k\rangle \in {\cal D}$, form a core ${\cal
C}({\cal L'})$ for the generator ${\cal L}'\equiv -i[H,.]+{\cal L}$, i.e., for any $\rho$ in
the domain of ${\cal L}'$ there exists a sequence $\rho_n\in{\cal
C}({\cal L}')$ such that $\rho_n \to \rho $, and ${\cal L}'(\rho_n) \to
{\cal L}'(\rho) $;\\

It follows from condition a) that the possibly infinite sums $\sum F_{\alpha}
F_{\alpha}^{\dagger}\rho^2$ and $\sum F_{\alpha}^{\dagger } \rho
F_{\alpha}\rho$ converge for all $\rho\in {\cal C}({\cal
  L}')$. Therefore the expression on the RHS of (\ref{eq:pdot1}) is
meaningful for all $\rho=\rho^{\dagger }\in {\cal C}({\cal L}')$ and
can be transformed into the form 
\begin{equation}
-\frac{1}{2}\sum_{\alpha }\mathrm{Tr}[(\rho F_{\alpha }- F_{\alpha}\rho)^{\dagger}(\rho F_{\alpha }- F_{\alpha}\rho)]
+\mathrm{Tr}(\rho ^{2}[F_{\alpha },F_{\alpha }^{\dagger }]).
\label{eq:pdot2}
\end{equation}
The first term is evidently negative. Then, due to
Eq.~(\ref{eq:condition}), this leads to $\dot{p}\leq 0$ first for all
$\rho\in {\cal C}({\cal L}')$, and then by condition b) for all $\rho$
in the domain of ${\cal L}'$ \cite{proof}.

One should notice that in the finite-dimensional case condition~(\ref{eq:condition}) is equivalent to 
$\sum[F_{\alpha }^{\dagger },F_{\alpha }]=0$.

\subsection{Proof using the Dissipativity Relation}

Lindblad uses properties of C$^{\ast }$-algebras to prove (Eq.~(3.2)
in \cite{Lindblad:76}) the general \textquotedblleft dissipativity
relation\textquotedblright :
\begin{equation}
{\mathcal{L}}(A^{\dag }A)+A^{\dag }{\mathcal{L}}(I)A-{\mathcal{L}}(A^{\dag
})A-A^{\dag }{\mathcal{L}}(A)\geq 0.
\end{equation}
This relation is valid for bounded generators ${\cal L}$
(though it may be possible to extend it to the unbounded case).
Taking the trace and using $\mathrm{Tr}[{\mathcal{L}}(A^{\dag }A)]=0$, $
A=\rho =\rho ^{\dag }$ then yields
\begin{equation}
\dot{p}=\mathrm{Tr}[\rho ({\mathcal{L}}\rho )]+\mathrm{Tr}[({\mathcal{L}}
\rho )\rho ]\leq \mathrm{Tr}[\rho ^{2}{\mathcal{L}}(I)].
\end{equation}
To guarantee $\dot{p}\leq 0$ it is thus sufficient to require
$\mathrm{Tr} [\rho ^{2}{\mathcal{L}}(I)]\leq 0$ for all states $\rho$. Using Eq.~(\ref{eq:lind2}) yields ${\mathcal{L}}(I)=\sum_{\alpha
}[F_{\alpha },F_{\alpha }^{\dagger }]$, so that
\begin{equation}
\dot{p}\leq \mathrm{Tr}\{\rho ^{2}\sum_{\alpha }[F_{\alpha },F_{\alpha
}^{\dagger }]\}\leq 0.  \label{eq:pdot}
\end{equation}
Since $\rho ^{2}>0$, it follows that it suffices for $\mathcal{F}\equiv
\sum_{\alpha }[F_{\alpha },F_{\alpha }^{\dagger }]$ to be a negative
operator in order for the inequality to be satisfied \cite{proof}. We have thus proved that ${\mathcal{L}}
(I)=\sum_{\alpha }[F_{\alpha },F_{\alpha }^{\dagger }]\leq 0$, is
sufficient.

\subsection{Proof using the Schwarz Inequality}

We now give an alternative and more direct proof of sufficiency, which,
again, is valid only in the case of bounded Lindblad operators. Let $
X_{\alpha }=\rho F_{\alpha }$ and $Y_{\alpha }=\rho F_{\alpha }^{\dagger }$,
and use this, along with $\sum_{\alpha }F_{\alpha }^{\dagger }F_{\alpha
}=\sum_{\alpha }F_{\alpha }F_{\alpha }^{\dagger }-A$, where $A\leq 0$ and
bounded, to rewrite Eq.~(\ref{eq:pdot1}) as 
\begin{eqnarray}
\dot{p} &=&2\mathrm{Tr}(\rho {\mathcal{L}}(\rho ))  \notag \\
&=&\mathrm{Tr}(\rho \sum_{\alpha }\left( [F_{\alpha },\rho F_{\alpha
}^{\dagger }]+[F_{\alpha }\rho ,F_{\alpha }^{\dagger }]\right) )  \notag \\
&=&\sum_{\alpha }2\mathrm{Tr}[\rho F_{\alpha }\rho F_{\alpha }^{\dagger
}]-\sum_{\alpha }\mathrm{Tr}[\rho F_{\alpha }^{\dagger }F_{\alpha
  }\rho ] \notag \\
&& - \mathrm{Tr}[\sum_{\alpha }\rho (\sum_{\alpha }F_{\alpha }F_{\alpha
}^{\dagger }-A)\rho ]  \notag \\
&=&\sum_{\alpha }{2\mathrm{Tr}[X_{\alpha }Y_{\alpha }]-\mathrm{Tr}[Y_{\alpha
}Y_{\alpha }^{\dagger }]-\mathrm{Tr}[X_{\alpha }X_{\alpha }^{\dagger }]}+
\mathrm{Tr}[\rho ^{2}A] . \notag \\
\end{eqnarray}
We can now apply the the Schwarz inequality for
operators 
\begin{eqnarray}
|\mathrm{Tr}(X^{\dagger }Y)| &\leq& \lbrack \mathrm{Tr}(X^{\dagger }X)]^{1/2}[
    \mathrm{Tr}(Y^{\dagger }Y)]^{1/2} \notag \\
  &\leq& \frac{1}{2}[\mathrm{Tr}(X^{\dagger }X)+
\mathrm{Tr}(Y^{\dagger }Y)],
\end{eqnarray}
and use the fact that $\mathrm{Tr}[X_{\alpha }Y_{\alpha }]\geq 0$ \cite{proof2}
to yield 
\begin{equation}
{2\mathrm{Tr}[X_{\alpha }Y_{\alpha }]-\mathrm{Tr}[Y_{\alpha }Y_{\alpha
}^{\dagger }]-\mathrm{Tr}[X_{\alpha }X_{\alpha }^{\dagger }]}\leq 0.
\end{equation}
Additionally, $\mathrm{Tr}[\rho ^{2}A]\leq 0$ (since $\rho ^{2}>0$ and $
A\leq 0$ by assumption), which completes the proof that $\dot{p}\leq 0$.

\section{Necessity: a condition for finite-dimensional Hilbert spaces}

\label{nece}

We would now like to derive a necessary condition on the Lindblad operators $
F_{\alpha }$ so that $\dot{p}\leq 0$ holds for all $\rho $. Our starting
point is again Eq.~(\ref{eq:pdot1}), which is valid only for bounded $
F_{\alpha }$: \ 
\begin{equation}
\dot{p}=\sum_{\alpha }\mathrm{Tr}[\rho F_{\alpha }\rho F_{\alpha }^{\dagger
}]-\mathrm{Tr}[\rho ^{2}F_{\alpha }^{\dagger }F_{\alpha }]\leq 0.  \label{EQ}
\end{equation}
We now restrict ourselves to the case of finite-dimensional Hilbert spaces.
The inequality (\ref{EQ}) must hold in particular for states $\rho $ which
are close to the fully mixed state, i.e., $\rho =(I\pm \varepsilon A)/
\mathrm{Tr}[I\pm \varepsilon A]$, where $0<\varepsilon \ll 1$ and $A=A^{\dag
}$, $\|A\|\leq 1$ and otherwise arbitrary. Let us find the constraint that must be obeyed
by the $F_{\alpha }$ so that (\ref{EQ}) is true for such states. This will
be a necessary condition on the $F_{\alpha }$.

Inserting this $\rho $ into the inequality~(\ref{EQ}) yields:
\begin{eqnarray}
0 &\geq &\sum_{\alpha }\mathrm{Tr}[(I\pm \varepsilon A)F_{\alpha }(I\pm
  \varepsilon A)F_{\alpha }^{\dagger }] \notag \\
&& -\mathrm{Tr}[(I\pm \varepsilon
A)^{2}F_{\alpha }^{\dagger }F_{\alpha }]  \notag \\
&=&\sum_{\alpha }\mathrm{Tr}[F_{\alpha }F_{\alpha }^{\dagger }\pm
\varepsilon AF_{\alpha }F_{\alpha }^{\dagger }\pm \varepsilon F_{\alpha
}AF_{\alpha }^{\dagger }] \notag \\
&& -\mathrm{Tr}[F_{\alpha }^{\dagger }F_{\alpha }\pm
2\varepsilon AF_{\alpha }^{\dagger }F_{\alpha }]+O(\varepsilon ^{2})  \notag
\\
&=&\sum_{\alpha }\mathrm{Tr}[F_{\alpha },F_{\alpha }^{\dagger }]\pm
\varepsilon \mathrm{Tr}[AF_{\alpha }F_{\alpha }^{\dagger }+F_{\alpha
  }AF_{\alpha }^{\dagger }-2AF_{\alpha }^{\dagger }F_{\alpha }] \notag \\
&& +O(\varepsilon^{2}).
\label{eq:eps}
\end{eqnarray}
The term $\mathrm{Tr}[F_{\alpha },F_{\alpha }^{\dagger }]$
vanishes, and the second term in the last line becomes $\mathrm{Tr}
(A\sum_{\alpha }[F_{\alpha },F_{\alpha }^{\dagger }])$. We may then divide
by $\varepsilon $ and take the limit $\varepsilon \rightarrow 0$, which
yields:

\begin{equation}
\pm \mathrm{Tr}(A\sum_{\alpha }[F_{\alpha },F_{\alpha }^{\dagger }])\leq 0.
\end{equation}
Since $A$ is arbitrary this result can only be true for all $A$ if $
\sum_{\alpha }[F_{\alpha },F_{\alpha }^{\dagger }]=0$. This is exactly the
unitality condition. We have thus proved Theorem~\ref{th:2}.

However, in the infinite-dimensional case the above argument fails since then
in general $\mathrm{Tr}F_{\alpha }F_{\alpha }^{\dagger }\neq \mathrm{Tr}
F_{\alpha }^{\dagger }F_{\alpha }$, or the trace may not even be defined.
Indeed, take a (non-invertible) isometry $V$ satisfying $V^{\dagger}V
= I$ and $VV^{\dagger} = P$ (a projector).
A physical example is given by the bosonic amplitude-raising semigroup (single Lindblad
operator): $F=a^{\dagger }=$the bosonic creation operator. Then ${\mathcal{L}
}(I)=[F,F^{\dagger }]=[a^{\dagger },a]=-I$, so the semigroup is non-unital.
Yet, ${\mathcal{L}}(I)<0$, so that by Theorem~\ref{th:1} (sufficiency) we
know that this is a purity-decreasing semigroup. Another example, where ${
\mathcal{L}}(I)\neq cI$ ($c$ a constant), yet ${\mathcal{L}}(I)\leq 0$, is
the case $F=a^{\dagger }aa^{\dagger }$ (this example is not even
trace-preserving). We thus have:

\begin{mycorollary}
In the infinite-dimensional case the purity may be
strictly decreasing without the Lindblad generator being unital.
\end{mycorollary}

An example of a semigroup that does not satisfy ${\mathcal{L}}(I)\leq 0$ is
the bosonic amplitude-damping semigroup, $F_{\alpha }=a$. For this semigroup ${
\mathcal{L}}(I)=[F_{\alpha },F_{\alpha }^{\dagger }]=[a,a^{\dagger }]=+I$.
Thus, this semigroup is not in general purity-decreasing. Indeed, amplitude
damping will in general purify a state $\rho $ by taking it to the ground
state $|0\rangle \langle 0|$, which is pure.

\section{Conclusions}

\label{conclusions}

In this work we have provided necessary and sufficient conditions for
Markovian open-system dynamics to be strictly purity-decreasing. These
conditions are summarized in Theorems~\ref{th:1},\ref{th:2}. It turns out
that the well-known result that unital semigroups are purity-decreasing is a
complete characterization (i.e., the condition is both necessary and
sufficient) for \emph{finite-dimensional}
Hilbert spaces. However, in the infinite-dimensional case it is
possible for a semigroup to be strictly purity-decreasing without being
unital. A simple example thereof is the bosonic amplitude-raising semigroup.
We leave as an open question the problem of finding a necessary condition in
the case of unbounded generators.

\begin{acknowledgments}
D.A.L. acknowledges the Sloan Foundation for a Research Fellowship, and
NSERC and PREA for financial support. R.A. acknowledges financial support
from the Polish Ministry of Scientific Research under Grant PBZ-Min-008/P03/03.
\end{acknowledgments}


\begin{thebibliography}{99}
\bibitem{comment-conds}
In order for Eq.~(\ref{eq:Lind}) to be meaningful even in the case of
unbounded operators we must define the domain of ${\cal L}' \equiv
-i[H,.] + {\cal L}$. This is assured by the stated conditions a) and
b). In particular, one starts with a ``predomain'' for which one
can explicitly define ${\cal L}'\rho$.  Such a ``predomain'' is given
in b), while a) is necessary to define the action of ${\cal L}'$ on an
appropriate joint domain for all operators involved.
\bibitem{proof}
  Here is a proof: $\rho^2 = \sum_j \lambda_j^2 |j\rangle \langle j|$
  in the basis in which $\rho$ is diagonal. As $\sum_\alpha [F_\alpha,F_\alpha^\dagger] <0$ means that for all vectors $|f\rangle$, the
  inner product $\langle f,\sum_\alpha [F_\alpha,F_\alpha^\dagger] f\rangle < 0$, obviously
  \begin{equation*}
{\rm Tr}(\rho^2 \sum_\alpha
  [F_\alpha,F_\alpha^\dagger]) = \sum_j \lambda_j^2 \langle j, \sum_\alpha
  [F_\alpha,F_\alpha^\dagger] j\rangle < 0.
\end{equation*}  
This calculation makes sense even for unbounded $\sum_\alpha
[F_\alpha,F_\alpha^\dagger]$ defined as a
quadratic form, provided the vectors $|j\rangle$  belong to the domain
of the quadratic form.
\bibitem{proof2}
Here is a proof:
  \begin{equation*}
{\mathrm{Tr}[X_{\alpha }Y_{\alpha }]=\mathrm{Tr}[\rho F_{\alpha }\rho
F_{\alpha }^{\dag }]}={\mathrm{Tr}[\rho }_{D}{F_{\alpha }^{^{\prime }}\rho }
_{D}{F_{\alpha }^{^{\prime }\dag }]} ,
\end{equation*}
where $U$ is the unitary matrix that diagonalizes $\rho $, $\rho _{D}={{U}
^{\dag }\rho U}$ is diagonal with eigenvalues $\lambda _{i}\geq 0$, and ${
F_{\alpha }^{^{\prime }}=U}^{\dag }{F_{\alpha }U}$. Explicitly evaluating
the trace yields:
\begin{eqnarray*}
{\mathrm{Tr}[\rho }_{D}{F_{\alpha }^{^{\prime }}\rho }_{D}{F_{\alpha
}^{^{\prime }\dag }]} &{=}&\sum_{ijkl}\lambda _{i}\delta _{ij}({F_{\alpha
}^{^{\prime }})}_{jk}\lambda _{k}\delta _{kl}({F_{\alpha }^{^{\prime }\dag })
}_{li} \\
&=&\sum_{ik}\lambda _{i}\lambda _{k}|({F_{\alpha }^{^{\prime }})}
_{ik}|^{2}\geq 0\text{.}
\end{eqnarray*}
\end{thebibliography}
\end{document}